\documentstyle[preprint,eqsecnum,aps]{revtex}

\newcommand{\be}{\begin{equation}}
\newcommand{\ee}{\end{equation}}
\newcommand{\bea}{\begin{eqnarray}}
\newcommand{\eea}{\end{eqnarray}}

\begin{document}

\input FEYNMAN
\preprint{HD-TVP-96-12}

\title{\vspace{1.5cm} Confinement in the Coulomb Gauge Model
      }

\author{ Th. Wilke and S.~P.~Klevansky }
\address{ Institut f\"ur Theoretische Physik, \\
Philosophenweg 16 u. 19, D-69120 Heidelberg, Germany}


\maketitle
\vspace{1cm}

\begin{abstract}
The Coulomb gauge model of QCD is studied with the introduction of
 a confining potential into the scalar part of the vector
potential.  Using a Green function formalism, we derive the self-energy for
this model, which has both scalar and vector parts, $\Sigma^S(p)$ and
$\Sigma^V(p)$.   A rotation of these variables leads to the so-called 
 gap and energy
equations. 
 We then analyse the divergence structure of these  equations.  
  As this depends explicitly on the form of potential, we give as examples
both the 
 linear plus
Coulomb and quadratically confining potentials.   The nature of the
confining single particle Green function is investigated, and shown to be
divergent due to the infrared singularities caused by the confining
potential.   Solutions to the gap equation for the simpler case of quadratic 
confinement are found both semi-analytically and numerically.   At finite
temperatures, the coupled set of equations are solved numerically in
two decoupling approximations.   Although chiral symmetry is found only
to be exactly restored as $T\rightarrow\infty$, the chiral condensate
displays a steep drop over a somewhat small temperature range. 

\end{abstract}

\clearpage

\section{Introduction}

From a phenomenological and experimental viewpoint, two aspects of
quantum chromodynamics (QCD) are believed to be fundamental, (a)
that chiral symmetry is spontaneously broken in the ground state and
(b) that partons are confined.    Both chiral symmetry and the phenomenon
of confinement underly a common speculation, viz. that each is believed
to give rise to a phase transition at a characteristic temperature or
density $T_\chi(T_c)$ and $\rho_\chi(\rho_c)$ respectively.    As a side
remark, one may note that these two phase transitions have a very different
character - the chiral order parameter is zero in the ordered phase, and
non-zero in the broken phase, in character with the notions of condensed
matter physics, while the confinement transition has an inverted structure:  
the non-zero value of the order parameter, or Polyakov line in this case,
usually characterizes the deconfined or ordered phase.   The current belief,
stemming
from lattice gauge simulations, is that, for QCD including quarks, the
critical temperatures coincide \cite{karsch}.   This feature is not understood on a firm 
basis, and is merely an empirical observation.

Phenomenological studies over the last decade have predominantly
investigated the chiral symmetry aspect of QCD.     For example, the 
Nambu--Jona-Lasinio (NJL) model \cite{nambu,reviews}  has enjoyed tremendous success, due to its
mathematical tractability and ease of use.   Calculations of the low-energy
mesonic and baryonic sectors demonstrate the well-known fact that chiral
symmetry is broken at $T=0$ in nature.    From a theoretical point of view,
one can study the model at finite temperatures and densities, but it is
clear that the thermodynamical potential and associated bulk quantities will
be
 dominated by inadmissable quark degrees of freedom in an intermediate
temperature region where they should  not exist at all \cite{pf}.   One conceivable
method of removing this problem is to start with a confining model in the
first place.   This in itself presents many new difficulties that are not 
inherent in the non-confining NJL model.   The fundamental theoretical 
questions that must be answered are (i) what is the analytic structure of
a Green function that describes a confined particle \cite{gribov},
 and (ii) what is the
role and connection of chiral symmetry, if any, with confinement and
the deconfining transition.  In addition, such a model must be
required to reproduce the excellent agreement of non-confining chiral models
with experiment, which in itself, is no mean task due to the added 
complexity.  
Yet a further complication is that the nature of an effective deconfinement 
potential is not precisely known.   Bag model calculations suggest that
confinement should be introduced in a scalar potential \cite{bag}, while
 Coulomb gauge
QCD naturally favors vector confinement.   More detailed study is required
to clarify this issue fully.
 Finally, even the understanding of these features still 
leaves the question open as to how the model should {\it de}confine
dynamically.

Several authors have attempted to address the problem that is associated
with (i) above, i.e. of the analytic structure of a Green function
 for a confined quark \cite{gribov,efimov}.
Because of the various problems that are connected with these attempts
(e.g. non-causality in the case of \cite{krewald}, and so on), this question is still
not completely settled.    In this paper, we do not make any assumptions about the
form of the single quark Green function.     Rather, we choose to 
study a model Lagrangian that contains a confining potential, and 
directly compute the Green function, assuming that it satisfies a 
Dyson-like equation.    The feature of confinement should reflect itself
{\it automatically} in the Green function and our task is to make this
evident. 
Choosing a form for a confining interaction is
not unique however.   We have chosen to study the Coulomb gauge model
\cite{finger}  that employs vector confinement 
as a starting point,  simply because this model has been
studied by several authors previously
 \cite{finger,adler,math,kle,alk,alk2,alk3,ley,ley2}.
    As such, this work is not entirely
new:  however, it unites previously unrelated work, and presents this
model in a conceptual fashion that is different from most interpretations.
Differences in the existing literature are clarified.   The temperature
dependence in this model is also examined.  
Here there are two different approaches that are possible, either that of Green
functions in the real or imaginary time formalism, see for example
\cite{kle},
or via a variational principle that is applied to the free energy
\cite{ley2}.   We
use the former approach - one can explicitly see that it leads to equations
that are entirely  equivalent to  those obtained in the latter approach.
Here there has been some conflict in the literature.    The gap equation
that
we have derived in this fashion
 corresponds to that derived by Ref.\cite{ley2} on varying the free
energy, but is in conflict with that of Ref.\cite{math} who disregard the
temperature dependence of the quasiparticle energy, as well as that of Ref.
\cite{alk} who simply write down  a different gap equation.   Further 
conflict arises also  in the interpretation of the quasiparticle energy.
In this paper, we utilize the knowledge gained from the invariance structure
at zero temperature, $T=0$,
 and implement a similar procedure at finite temperatures.   Central to our
approach is the appearance of physical quasiparticle energies that are 
temperature dependent, as is the case in the BCS theory of superconductivity
for low temperatures.     It is essentially this temperature dependence
which
is capable of driving a phase transition, and which is studied here. This is
in  
apposition to models that introduce an effective temperature dependence in
the coupling strength, see, for example Ref.\cite{alk}, which is difficult
to deal with consistently from a  thermodynamical point of view.  In the
case studied here,
 we keep the coupling constant fixed and examine
the consequences of the temperature dependent quasiparticle energies.
 Numerical solutions
are given for the coupled set of equations that form the gap equation in
this case.   It is interesting to note that although in part only
approximate solutions can be given, the model dependence of $\langle
\bar\psi\psi\rangle$
closely resembles that of the lattice simulations \cite{laermann}.
   Chiral symmetry, is
however, not restored exactly.
 It would thus be of
great interest in the future to have lattice simulations that could
quantify the range of the drop in temperature of the
chiral order parameter.

Finally, one should note that the Coulomb gauge model, while appearing
intuitively
physical, is by definition no longer manifestly gauge covariant, as is
the non-confining NJL model.   However, since any potential introduced into
the NJL model would also be expected to be instantaneous, this covariance
is unavoidably lost.   Thus a study of the Coulomb gauge model should 
provide many features that should be generic to four-fermion interactions
and should be similarly handled in, for example, an extended NJL model
that includes confinement.

We point out that currently there has been a surge of interest in the
problem of confinement.   For different approaches, we
 refer the interested reader also to the works
of Refs.\cite{munz,roberts,gross,maris,rho}.   This list of references is,
however, not exhaustive.

This paper is organized as follows.   In Section II, the general formalism
is developed for $T=0$ for an arbitrary potential.   The form of the single
particle propagator is given and the distinction between observables and 
non-observables  
is made according to the conditions of ultraviolet renormalizability as well
as infrared finiteness.   These
    divergences are discussed for two particular
types of potentials, the Coulomb plus linear rising potential and the 
quadratically confining potential.   In Section III, the solutions of the
gap equation for a quadratically confining potential are discussed in this
framework.    In Section IV, the extension to finite temperatures is given,
and numerical solutions for the condensate and dynamical mass is given.
We summarize and conclude in Section V.

\section{Formalism for $T=0$} \label{form}

\subsection{Self-consistent self-energy}

Our starting point is the Coulomb gauge QCD Hamiltonian, given by
\begin{eqnarray}
\hat H & = & \hat H_0 + \hat H_1 \nonumber \\
 & = & \int d^3 x \hat \psi^\dagger ({\vec x})(-i\vec
\alpha\cdot\vec\nabla)
\hat \psi({\vec x})  \\
&+& \sum_{a=1}^8\int d^3xd^3y \hat\psi^\dagger({\vec x})\frac{\lambda_a}{2}
\hat\psi(\vec x) V({\vec x} - {\vec y})\hat \psi^\dagger({\vec
y})\frac {\lambda_a}{2}\hat\psi({\vec y}), \nonumber
\end{eqnarray}
where  $\lambda_a$ are the standard Gell-Mann matrices, and $V({\vec x})$
is an arbitrary potential at this point, and which will be considered to be
confining.    No current quark masses are considered.      

In general, the self-consistent self-energy  $\Sigma(p)$ associated with
the interaction is a function of momentum.
Under the assumption that the quark propagator is diagonal and independent
of
color, $S(k)_{c,c'}= S(k)\delta_{c,c'}$, one has the mathematical
consequence that the Hartree term vanishes.   Thus  
the leading non-vanishing
contribution to the self-energy, calculated to lowest order in the
interaction strength follows from the Fock diagram that is
depicted in Fig.1.  One finds
\begin{equation}
\Sigma(\vec p) = i\sum_{a=1}^8\int\frac{d^4k}{(2\pi)^4}V({\vec p}-{\vec
k})\gamma_0\frac{\lambda_a}2 S(k) \gamma_0\frac{\lambda_a}2,
\label{sig}
\end{equation}
where $S(k)$ is the self-consistently evaluated one particle fermion Green
function, of form
\begin{equation}
S(p)^{-1} = \not p - \Sigma(p).
\label{es}
\end{equation}
Note that this structure is particular to this form of the interaction, and,
as such, may not be taken as being generic to confinement {\it per se}.
Lorentz symmetry and invariance under a parity transformation enable one to 
decompose $\Sigma(p)$ into a scalar and vector term,
\begin{equation}
\Sigma(\vec p) = \Sigma^S(|\vec p|) + \vec \gamma\cdot\hat p \Sigma^V(|\vec p|),
\label{dec}
\end{equation}
where $\hat p  = \vec p/|\vec p|$ is the unit vector.   Inserting the  
{\it ansatz} Eq.(\ref{dec}) into Eq.(\ref{es}) leads to the explicit form
\begin{equation}
S(p_0,\vec p) = \frac{p_0\gamma_0 - \vec\gamma\cdot \hat p(p+\Sigma^V(p)) + 
\Sigma^S(p)}{p_0^2 - [(p+\Sigma^V(p))^2 + (\Sigma^S(p))^2]}
\label{prop}
\end{equation}
for the propagator, which, when inserted into Eq.(\ref{sig}), leads to the
coupled set of non-linear integral equations,
\begin{eqnarray}
\Sigma^S(p) &=& \frac 2 3\int\frac{d^3k}{(2\pi)^3} V(\vec p - \vec k)
\frac{\Sigma^S(k)}{[(k+\Sigma^V(k))^2 + (\Sigma^S(k))^2]^{1/2}} \\
\Sigma^V(p) &=& \frac 2 3 \int\frac{d^3k}{(2\pi)^3} V(\vec p - \vec k) \frac
{(k+\Sigma^V(k)) \hat p\cdot\hat k}{[(k+\Sigma^V(k))^2 +
(\Sigma^S(k))^2]^{1/2}} 
\label{coupled}
\end{eqnarray}
for the unknown scalar and vector self-energies.   The form of
Eq.(\ref{prop}) and also (\ref{coupled}) suggests a reparametrization of 
$\Sigma^S(p)$ and $\Sigma^V(p)$ as
\begin{mathletters}\label{ep}
\begin{eqnarray}
\Sigma^S(p) & = & E(p)\sin \phi(p) \\
 \mbox{and} \quad p + \Sigma^V(p) & = & E(p)\cos \phi(p) 
\end{eqnarray} \end{mathletters}
introducing the variable $E(p)$ and the angular variable $\phi(p)$.   It
can be seen that the variable $E(p)$ plays the role of an energy, by 
writing the  propagator with respect to the new variables:
\begin{equation}
S(p_0,\vec p) = \frac{\gamma^0p_0 - \vec\gamma\cdot\hat p E(p)\cos (\phi(p)) 
+E(p) \sin(\phi(p))}{p_0^2 - E(p)^2}, 
\label{s2}
\end{equation}
while the Eqs.(\ref{ep}) are replaced by
\begin{mathletters}\label{ep2}
\begin{eqnarray}
E(p)\sin(\phi(p)) & = & \frac 2 3 \int\frac{d^3k}{(2\pi)^3} V(\vec p - 
\vec k) \sin(\phi(k)) \\
E(p)\cos (\phi(p)) - p & = & \frac 2 3 \int\frac{d^3k}{(2\pi)^3}V(\vec p
- \vec k) \cos (\phi(k))\hat p\cdot \hat k. 
\end{eqnarray}
\end{mathletters}
A simple manipulation of Eqs.(\ref{ep2}) can be made to decouple the 
unknowns $E(p)$ and $\phi(p)$, leading to the result
\begin{equation}
p\sin(\phi(p)) = \frac2 3\int\frac{d^3k}{(2\pi)^3}V(\vec p - \vec
k)[\sin(\phi(k))\cos(\phi(p)) - \hat p\cdot\hat
k\cos(\phi(k))\sin(\phi(p))],
\label{gapeqn}
\end{equation}
that determines $\phi(p)$ solely.  One may note that the chiral condensate
can be expressed purely in terms of $\phi(p)$ as
\begin{equation}
\langle\bar\psi\psi\rangle = -\frac 3{\pi^2}\int_0^\infty dk k^2
\sin(\phi(k)),
\label{cond}
\end{equation}
from which it is clear that $\phi(p)$ is the order parameter characterizing
a non-trivial ground state.   The determining equation for this function,
Eq.(\ref{gapeqn}) is thus the
{\it gap equation} \  for an arbitrary potential in this model. 
Assuming that the symmetry breaking is maximal for the particle at rest, one
may impose the boundary conditions
\begin{equation}
\lim_{p\rightarrow 0}\phi(p) = \frac {\pi}2,
\label{bc1}
\end{equation}
while for large $p$,
\begin{equation}
\lim_{p\rightarrow\infty}\phi(p) \rightarrow 0.
\label{bc2}
\end{equation}
This completely specifies the problem.

\subsection{Confinement, physical  and unphysical variables}

While the formal manipulations of the previous section are simple, some
interpretation of the derived equations is required for a potential that is
confining.   In particular, the simple form of the propagator in
Eq.(\ref{s2}) might lead one to believe that this function has singularities
in the complex plane.   In fact, this is not so.  Closer inspection of the 
defining relation for $E(p)$ in Eq.(\ref{ep2}) indicate that it is related
to the integral of the function $V(\vec p - \vec k)$, which, for a confining
potential, is infrared singular.    Thus $E(p)$ may {\it not} be interpreted
as the physical energy of a quark.   Only variables that are found to be
infrared finite (and, if necessary,
 ultraviolet renormalizable) can be regarded in a
physical sense \cite{adler}.

We thus assert that:
(i) $E(p)$ is, in general, divergent, and thus necessitates a redefinition
of the physical energy in order to extract a physical quasiparticle mass, and
(ii) $\phi(p)$ is a physical quantity for both linear and quadratic
confining potentials, as the gap equation that it satisfies is infrared 
finite.  In the following two subsections, we examine the
possible divergences that may be associated with the infrared, $\vec k
\rightarrow 0$ and ultraviolet $k\rightarrow\infty$ behavior of the
gap and energy equations respectively.

\subsubsection{Ultraviolet divergences}

\paragraph{Gap Equation}   For $p\rightarrow\infty$,
$\sin(\phi(p))\rightarrow 0$ and $\cos(\phi(p))\rightarrow 1$.   The first
term
 of the gap equation, Eq.(\ref{gapeqn}) is thus ultraviolet convergent.   A
divergence could however arise from the second term, depending on the form
of the potential.   For a Coulomb-like potential of the form $V(r) \sim
-1/r$ or $V(q)\sim 1/q^2$, a divergence going as
\begin{equation}
\int d^3k \frac 1{k^2} \sim \int dk k^2\frac 1{k^2}
\label{s3}
\end{equation}
occurs.
In general, the gap equation is ultraviolet divergent for potentials that
fall off more weakly than the Coulomb potential.  In order to quantify this
statement, consider a potential of the general form $V=V(|\vec p - \vec
k|^2)$.   The leading $k$ dependence is obtained from the Taylor series 
of this function about $\vec p=0$, as
\begin{equation}
V(|\vec p - \vec k|^2) = V(k^2 ) - 2\vec p\cdot \vec k\frac{\partial V(k^2)}
{\partial (k^2)}+ \dots
\label{s4}
\end{equation}
and inserting this into the divergent term of the gap equation
\begin{equation}
\frac{2}{3}\int\frac{d^3k}{(2\pi)^3} V(|\vec p- \vec k|^2) \hat p\cdot \hat
k\sin(\phi(p)),
\label{s5}
\end{equation}
one sees, after performing the angular integration that the leading
 possibly divergent contribution 
 goes as
\begin{equation}
\int d^3k \frac{\partial V(k^2)}{\partial(k^2)} k.
\label{s6}
\end{equation}
For a confining potential that has the limiting behavior
\begin{equation}
V(q^2)\stackrel{q\rightarrow\infty}{\sim} (q^2)^{-\alpha}, \quad \alpha>0,
\label{s7}
\end{equation}
Eq.(\ref{s6}) can be evaluated to give the result $k^{-2\alpha+2}/(-2\alpha + 2)$,
which
thus exhibits a divergence for $\alpha\le 1$.    Denoting 
\begin{equation}
Z-1 = -\frac4 9\int\frac{d^3k}{(2\pi)^3}\frac{\partial
V(k^2)}{\partial(k^2)} k,
\label{s8}
\end{equation}
one finds that the divergent part of the gap equation, i.e. Eq.(\ref{s5}), is
of the form
\begin{equation}
\mbox{divergence} = (Z-1)p\sin(\phi(p)).
\label{s9}
\end{equation}
This, however, has exactly the form that appears on the left hand side of
the gap equation, and thus any ultraviolet divergences can be simply
compensated by introducing an appropriate counter term into the Hamiltonian,
in this case,
\begin{equation}
\Delta \hat H = (Z-1) \int d^3 x\hat \psi^\dagger(\vec x)(-i\vec
\alpha\cdot\nabla)\hat\psi(\vec x).
\end{equation}
With this renormalization prescription, ultraviolet divergences can 
be unambiguously dealt with.

\paragraph{Energy equation}   Since $\phi(k)\rightarrow 0$ when 
$k\rightarrow \infty$, it follows from Eq.(\ref{ep2}) that $E(p)$ is 
ultraviolet convergent.

\subsubsection{Infrared divergences}

\paragraph{Gap equation}
By performing an expansion of the bracketed term in the integrand of
Eq.(\ref{gapeqn}) about $p=k$, i.e.
\begin{eqnarray}
\sin(\phi(k))\cos(\phi(p))& - &\hat p\cdot\hat k\cos(\phi(k))\sin(\phi(p))
\nonumber \\
& = & (\vec p - \vec k)\cdot \hat p\phi'(p) + O((\vec p - \vec k)^2),
\end{eqnarray}
and shifting the variable $\vec q = \vec p - \vec k$ in the integrand of
the gap equation  one may isolate the possible divergence as being
\begin{equation}
\mbox{divergence} \sim \frac 23\int\frac{d^3q}{(2\pi)^3} V(\vec q)O(q^2).
\label{s10}
\end{equation}
Thus, for a linear confining potential, $V(r)\sim r$ or $V(q)\sim 1/q^4$,
this
term takes the form
\begin{equation}
\int d^3q \frac 1{q^4} O(q^2) \sim \int dq q^2\frac 1{q^4} O(q^2) \sim \int
dq,
\label{s11}
\end{equation}
which is infrared finite.    Thus the gap equation is non-singular in the
infrared in this case.
     This can also be verified to be true for a potential that
is quadratically confining, i.e. $V(r) \sim r^2$.

\paragraph{Energy equation} Eq.(\ref{ep2}) can be used to determine the
quasiparticle energy,
\begin{equation}
E(p) = \frac23\int\frac{d^3k}{(2\pi)^3} V(|\vec p - \vec k|)
\frac{\sin(\phi(k))}{\sin(\phi(p))} = \frac 23\int\frac{d^3k}{(2\pi)^3}
V(k)\frac{\sin(\phi(|\vec p - \vec k|))}{\sin(\phi(p))}.
\label{s12}
\end{equation}
The divergence structure in the infrared follows on
expanding the sine
function about $\vec k = \vec p$.    The leading term is
\begin{eqnarray}
\mbox{divergent part of}\  E(p) &\sim& \frac 23\int \frac{d^3k}{(2\pi)^3} V(k)
\nonumber \\
&\sim& \mbox{divergent part of}\  E(0),
\label{s13}
\end{eqnarray}
the latter statement of which can be seen by inspection.  It is thus evident
that the energy {\it difference}
\begin{equation}
E(p) - E(0)
\label{s14}
\end{equation}
is infrared finite.  

One may note that
  the subtraction of the infinite zero point energy
corresponds to shifting the potential by an infinite constant 
\begin{equation}
V(\vec r) \rightarrow V(\vec r) - \frac 23\int\frac{d^3k}{(2\pi)^3}V(k)
\label{s15}
\end{equation}
or
\begin{equation}
V(\vec q) \rightarrow V(\vec q) - (2\pi)^3\delta^{(3)}(\vec q)\frac 23\int\frac
{d^3k}{(2\pi)^3}V(k).
\label{s16}
\end{equation}
Thus the content of this section on infrared divergences can also be viewed in
the context that physical observables are invariant under shifts of the 
potential by a constant.   This is immediately obvious for the gap equation,
Eq.(\ref{gapeqn}), but not for $E(p)$ alone as in Eq.(\ref{s12}).
Under the general shift 
\begin{equation}
V(\vec r) \rightarrow V(\vec r) + U,
\label{s17}
\end{equation}
it follows that $E(p)\rightarrow E(p) + 2 U/3$.  Consequently, the
determination of $E(0)$ does not provide a unique determination of the
quasiparticle mass.    Following Ref.\cite{adler}, one may make an 
identification of the quasiparticle mass by rewriting the single particle
Green function of Eq.(\ref{prop}) as
\begin{equation}
S(p_0,\vec p) = \frac{R_+(\vec p)}{p_0 + (E(p)-E(0)) + E(0))} +
 \frac{R_-(\vec p)}{p_0-(E(p)-E(0))-E(0))},
\label{s18}
\end{equation}
where the residues
\begin{equation}
R_\pm(\vec p) = \frac 12(\gamma_0 \mp \vec\gamma\cdot\hat p\cos(\phi(p))\pm
\sin(\phi(p)))
\label{res}
\end{equation}
are {\it unaffected} by a shift in the potential.   By comparing the residues
at low momenta with those of a free relativistic particle of mass
$m^*$, viz. $R_\pm^0 = \frac 12(\gamma^0\mp\vec \gamma\cdot\vec p/m^* \pm
1)$, one arrives at an unambiguous identification
\begin{equation}
\frac 1{m^*} = \lim_{p\rightarrow 0}\frac{\cos(\phi(p))}{p}
\label{mass}
\end{equation}
for the quasiparticle mass.

It is however physically more intuitive to introduce the {\it shifted} energy
\begin{equation}
\bar E(p) = E(p) - E(0) + m^*
\label{ebar}
\end{equation}
which has the desirable property $\bar E(0) = m^*$.    In order to 
ensure this property, the shift
\begin{equation}
\frac 23U = -E(0) + m^* 
\label{shift}
\end{equation}
must be invoked in the potential Eq.(\ref{s17}).

In concluding this section, we point out that both $m^*$ and the condensate
$\langle \bar\psi \psi\rangle$, calculated via Eqs.(\ref{mass}) and
(\ref{cond})
respectively, fullfil the requirements of being physical observables as they
are solely functions of the gap variable $\phi$.

\section{Solutions for quadratic confinement at $T=0$}

In this section, we illustrate the solution of the gap equation for a 
quadratically confining potential, since such a form for the potential
has
the advantage that the integral gap and energy equations can be reduced to 
differential forms, even although they are highly non-linear, and the
problem is a boundary value, not an initial value one.     Nevertheless, many
analytic techniques are at ones disposal, and in this case, precise
solutions can be found.    Note that this problem has been solved
approximately elsewhere \cite{ley},
 but not within the physical framework presented
here.   Here  we give a more general solution than that in Ref.\cite
{ley}. Since the method will also be applied to the problem at finite temperatures in
Section IV, we do this in some detail.

Our starting point is the attractive potential
\begin{equation}
V(\vec r) = -\lambda \vec r^2 + U,
\label{quad}
\end{equation}
with $\lambda>0$ controlling the effective strength of the potential, and 
$U$ determined via the condition Eq.(\ref{shift}). 
This potential is binding in the quark-antiquark channel.
  The Fourier transform
of the potential Eq.(\ref{quad}) is given as
\begin{equation}
V(\vec q) = \lambda(2\pi)^3\nabla^2_q\delta^{(3)}(\vec q) + U (2\pi)^3
\delta^{(3)}(\vec q).
\label{pot}
\end{equation}
Inserting this into the integral gap equation, Eq.(\ref{gapeqn}) and
performing an integration by parts, leads one to the differential form
\begin{equation}
p\sin(\phi(p)) = \frac 23\lambda\large(\phi'{}' (p) + \frac 2 p\phi'(p) + \frac 1
{p^2}\sin(2\phi(p))\large),
\label{gapeqnn}
\end{equation}
which is to be solved subject to the boundary conditions Eqs.(\ref{bc1}) and
(\ref{bc2}).   A semi-analytic solution to this problem is presented in the
following subsection.  This is later compared with an
 exact numerical solution.

\subsection*{Semi-analytic solution} 

The differential gap equation, written as
\begin{equation}
ap^3\sin(\phi(p)) = p^2\phi'{}'(p) + 2p\phi'(p) + \sin(2\phi(p))
\label{difge}
\end{equation}
with $a^{-1} = 4\lambda/3>0$ can be solved by the technique of asymptotic
matching \cite{bender}.  In this method, one identifies overlapping regions
in the
independent variable, for which exact analytic solutions can usually be
obtained.   On enforcing the boundary conditions, these solutions are
matched onto one another asymptotically in their common region of validity.
Four such regions can be identified, of which the second and fourth are 
 a limiting cases
of the third and are thus not independent.   These regions are
\begin{mathletters}\label{cases}
\begin{eqnarray}
(1)\  p \rightarrow 0 \quad&\Rightarrow&\quad  p^2\phi'{}'(p) + 2p\phi'(p) = -\sin(2\phi(p)) \\
(2)\   p \ \mbox{and} \ \phi(p) \ \mbox{small} \quad&\Rightarrow&\quad p^2\phi'{}'(p) +
2p\phi'(p) = -2\phi(p) \\
(3)\  \phi(p)\  \mbox{small for arbitrary } \ p \quad&\Rightarrow&\quad p^2\phi'{}'(p) +
2p\phi'(p) = (ap^3-2)\phi(p) \\
(4)\  \phi(p) \ \mbox{small and}\  p\rightarrow\infty\quad&\Rightarrow&\quad p^2\phi'{}'(p)
+2p\phi'(p) = ap^3\phi(p).
\end{eqnarray}
\end{mathletters}

Correspondingly,
the energy equation, in differential form, is given as
\begin{equation}
\bar E(p) = -\frac 1a\large(\phi'(p)^2 + \frac 2{p^2}\cos^2(\phi(p)\large) +
p\cos(\phi(p)) + \frac 2 3 U.
\label{en}
\end{equation}
Thus, once one has knowledge of $\phi(p)$, the energy may be immediately
determined.    

We now deal with each of the above cases in turn.

\paragraph*{case 1}:    The determining equation Eq.(\ref{cases}a) in
this case cannot be solved exactly due to non-linearities.    However, since
the boundary condition Eq.(\ref{bc1}) must be imposed, it is useful to write
$\phi(p) = \pi/2 + \epsilon(p)$ with $\epsilon(0)=0$.  One may insert this
{\it ansatz} into Eq.(\ref{cases}a) and solve the resulting equation exactly for
small $\epsilon$.   The non-divergent solution is found to be linear in
$p$ so that one may write, in leading order,
\begin{equation}
\phi(p) \simeq \frac\pi 2 + cp + \eta(p), \quad p\rightarrow 0, \quad \eta
(0)=0
\label{smp}
\end{equation}
where $c$ is as yet an undetermined constant.   Clearly this procedure can
be repeated by constructing the differential equation for $\eta(p)$ to 
obtain further corrections.   This leads to the higher order correction
$\eta(p) = bp^3 + O(p^4)$, with $b=(a-4c^3/3)/10$.   Note that the constant
$c$ in Eq.(\ref{smp}) plays an extremely important role.   According to
Eq.(\ref{mass}) and Eq.(\ref{smp}),
\begin{equation}
m^*{}^{-1} = -\phi'(0) = -c,
\label{conn}
\end{equation}
i.e. the initial slope of the gap angle is related to the inverse 
quasiparticle mass,   This determines $c$ to be negative, $c=-|c|$.

We now turn to the determination of $\bar E(p)$.   Using the approximation
Eq.(\ref{smp}) in Eq.(\ref{en}), one finds
\begin{equation}
\bar E(p) = -\frac {3c^2}{a} + [\frac {2c^4}{a} - 2c]p^2 + \frac 2 3 U, 
\quad p\rightarrow 0.
\label{sme}
\end{equation}
The criterion that $\bar E(0) = m^*$ now fixes $U$ to be
\begin{equation}
U=\frac 92 \frac{c^2}{a} + \frac 3 2\frac 1{|c|}.
\label{ueq}
\end{equation}
It is interesting to note that, in light of these comments, the dispersion
relation for $\bar E(p)$ for small momenta, from Eq.(3.8),
\begin{equation}
\bar E(p) = m^* + Bp^2, \quad p\rightarrow 0,
\label{dis}
\end{equation}
with $B= 2c^4/a - 2c>0$, is non-relativistic.

\paragraph*{case 2}:   In the region specified as (2), the determining
equation Eq.(\ref{cases}b) can be solved exactly.  The solution is
\begin{equation}
\phi_2(p) = \frac\gamma{\sqrt p} \cos(\frac{\sqrt7}2\ln(p) + \delta)
\label{two}
\end{equation}
where $\gamma$ and $\delta$ are two constants of integration.

\paragraph*{case 3}:   The determining equation over this region, i.e.
Eq.(\ref{cases}c) is a Bessel equation.    Selecting the solution that falls
off with momentum leads to the solution
\begin{equation}
\phi_3(p) = Bp^{-1/2} K_{i\sqrt7/3}(\frac23\sqrt ap^{3/2}),
\label{three}
\end{equation}
where $K_n(x)$ is the $n$th order modified Bessel function of the second
kind, and $B$ is an integration constant.

\paragraph*{case 4}: Eq(\ref{cases}d) is a limiting case of
Eq.(\ref{cases}c) and has as decaying solution, the modified Bessel function
\begin{equation}
\phi_4(p) = Cp^{-1/2} K_{1/3}(\frac 23\sqrt a p^{3/2}).
\label{four}
\end{equation}
For large $x$, $K_n(x)\sim \sqrt{\frac{\pi}{2x}}e^{-x}[1+O(1/x)]$, so that
\begin{equation}
\phi_4(p) \sim Ap^{-5/4} \exp(-\frac 2 3\sqrt a p^{3/2})
\label{fourlim}
\end{equation}
in the limit $p\rightarrow\infty$.

We now turn to the energy determination.  From Eq.(\ref{en}), it follows
that
\begin{equation}
\bar E(p) \stackrel{p\rightarrow\infty}{\sim} p.
\label{beh}
\end{equation}
Thus one recovers a relativistic dispersion relation at high momenta.

The next issue is to match all solutions that have been obtained piecewise
to construct one continuous solution over the entire momentum range.

\paragraph*{Asymptotic matching over the intervals.}   One may observe
that the Eqs.(\ref{cases}a) and (b) are scale invariant, i.e. invariant
under
the transformation $p\rightarrow |c|p$.   This enables one to determine the
constants $\gamma$ and $\delta$ for the solution (\ref{two}) without
actually knowing
$c=\phi'(0)$.   One determines the solution to Eq.(\ref{cases}a) with
initial values
\begin{equation}
\phi_1(0) = \frac \pi 2 \quad \mbox{and} \quad \phi_1{}'(0) = -1,
\label{co}
\end{equation}
and one considers this solution for $p\rightarrow\infty$.
This is matched onto the solution $\phi_2(p)$ from Eq.(\ref{two}) and the
associated $\phi'{}_2(p)$ in the regime $p\rightarrow \infty$ of the
$phi_1(p)$ equation.
This is possible since the solution of $\phi_2(p)$ can be matched to that
of $\phi_1(p)$ down to small momenta.  On the other hand, both
Eqs.(\ref{cases}a)
and (\ref{cases}b)  have the same behavior in the limit $p\rightarrow 0$.
One is thus able to extract the constants for the solution in region (2).

Numerically, one finds
\begin{eqnarray}
\gamma &=& 0.700460 a^{-1/6} \nonumber \\
\delta &=& 0.238503. 
\end{eqnarray}
  Now  the solution in region (2), $\phi_2(p)$, must be matched onto
the solution in region (3),  $\phi_3(p)$.   To
do this, one requires $\phi_3(p)$, for $p\rightarrow 0$.   Since
\begin{equation}
K_{i\nu} \stackrel{x\rightarrow0}{\sim}
 \frac12\Gamma(i\nu)(\frac12 x)^{-i\nu}
+ c.c.,
\end{equation}
one finds
\begin{eqnarray}
\phi_3(p) &\stackrel{p\rightarrow 0}{\sim}& Bp^{-1/2}|\Gamma(\frac{\sqrt{7}}3)|\frac 12
[\exp(i\frac{\sqrt{7}}3\ln(\frac{\sqrt{a}}3 p^{3/2})-i\rho) + c.c] \\
&=& Bp^{-1/2}|\Gamma(\frac{\sqrt{7}}3)|\cos(\frac{\sqrt{7}}2\ln(p) + \frac
{\sqrt{7}}2\ln(\frac{a^{1/3}}3) - \rho),
\label{p3}
\end{eqnarray}
where $\rho = \arg(\Gamma(\frac{i\sqrt{7}}3)) \simeq -1.877978$.    The
full, non-scaled solution in region (2),
\begin{equation}
\phi_2(p) = \frac{\gamma}{\sqrt{|c|p}}\cos(\frac{\sqrt{7}}2\ln(|c|p) + 
\delta)
\end{equation}
can be directly compared with Eq.(\ref{p3}), to give
\begin{equation}
\delta + \frac{\sqrt 7}{2} \ln|c| = \frac{\sqrt 7}2\ln(\frac{a^{1/3}}3) -
\rho + n\pi
\label{tt1}
\end{equation}
and
\begin{equation}
\frac{\gamma}{\sqrt{|c|}} = B|\Gamma(i\frac{\sqrt7}3)|.
\label{tt2}
\end{equation}
One may now extract the slope $|c|$ from Eq.(\ref{tt1}). Explicitly,  one has
\begin{equation}
|c| = \frac{a^{1/3}}3 \exp{\frac 2{\sqrt 7} (n\pi - \rho-\delta)}.
\label{tt3}
\end{equation}
where $n$ is the number of nodes occurring in the solution.   
In Table I, we list the node and the value for $c$ that were extracted
via this asymptotic matching procedure via expression (\ref{tt3}) 
in comparison with a trial exact
procedure in which one attempts to solve the full equation starting as near
to the origin as is  numerically permissible.

In Fig.2, we show the nodeless solution for the gap parameter.    The 
analytic solution in regions (1), (2) and (3) are indicated on the graph by
various symbols.One notes that the approximation is in excellent agreement
with the numerical result.   This procedure has proven extremely useful
both as a check but also in guiding the numerical solutions, since the 
gap equation is highly non-linear.
Figure 3 indicates the first three solutions for $\phi(p)$.   One notes that
the slope at the origin increases sharply with the number of nodes,
indicating that a succession of values of $m^*$
exist that {\it decrease} with the number of
nodes.

\paragraph*{Energy Density.}
In order to determine which  of the solutions, i.e. that with no node
present
or alternatively with  multiple nodes present 
should give rise to  the ground state of the system,
 one needs to evaluate the change in the  energy density that 
occurs in going from the normal vacuum to the condensed one.    The
physical solution to the gap equation 
is then determined as the solution that  lowers the
energy the most.     The vacuum energy density in the condensed phase, 
$W_{condensed}=\langle
cond|H|cond\rangle$ or in the normal phase,
$W_{normal}=\langle0|H|0\rangle$,
 can
be obtained directly from the appropriate 
single particle Green's function, via the
formula \cite{walecka}
\begin{equation}
W_{c,n} = -\frac i 2 \int\frac{d^4p}{(2\pi)^4} Tr[\gamma^0p_0 +
\vec\gamma\cdot \vec p] S_{c,n}(p_0,\vec p) e^{i\eta p_0}
\label{u1}
\end{equation}
where $c,n$ refer to the condensed or normal phases respectively.     This
 equation
is general and exact for a four-point or two-body interaction, such as
that under study here.   Explicitly inserting the
Green function, Eq.(\ref{s18}), for the condensed phase and the free Green
function for the normal phase, one finds the expression
\begin{equation}
\Delta W = W_c - W_n = \frac{3}{2\pi^2}\int_0^\infty dp
p^2[2p(1-\cos(\phi(p))) - \frac 1{ap^2}\sin^2\phi(p) + \frac
1{2a} \phi'{}^2(p)] 
\label{u2}
\end{equation}
for the change in energy density in moving from the normal to condensed 
vacua.   A numerical analysis indicates that $\Delta W$ is negative, and
that the nodeless solution, i.e. the solution with the smallest slope or
largest mass,  always leads to the lowest energy state.

\section{FINITE TEMPERATURE}

To date, there have been several studies of the finite temperature gap
equation.   These differ markedly in (a) the gap equation actually proposed
or derived and the subsequent interpretation of the variables, and (b) the
form of the potential chosen for studying this model.     In the first part
(a), there can of course be only one correct equation.   In analogy to the 
$T=0$ case, it can be formally derived by one of two equivalent methods, the
Green function approach, similar to that followed at $T=0$ in Section II of
this paper, or via a variational method that proceeds by constructing the
Gibbs free energy and minimizing it with respect to a variational parameter,
see Ref.\cite{ley2}.   In this section, we will take the former approach, since
the generalization from Section II is evidently simple.   We contrast this
result with the conflicting postulated gap equations of other authors
\cite{alk}
and interpret the quasiparticle energy that occurs along the
lines of Refs.\cite{adler} and \cite{kle}.

With regard to point (b) above, let us mention that the only attempt to date
of extracting a numerical solution to a finite temperature gap equation (not
that derived here) employs an {\it a priori} temperature dependent coupling
strength for a Coulomb-like potential \cite{alk}.   This, in itself, presents
difficulties in the construction of a consistent thermodynamic description.
In the case handled, the second order phase transition that was observed was
due to the coupling, that strongly modifies the infrared behavior of the
potential, rather than the inherent temperature dependence of the
quasiparticle energy, which was omitted in this description.

An additional difficulty manifests itself in the finite temperature
formalism with regard to the zero temperature one.   Since statistical
averages over a grand canonical ensemble are made, all states are included
in the averaging process.   However, physical states should be colorless.
Imposing this restriction leads to rather complex technicalities and a 
complex set of equations, from which it is not evident what the physics
actually is \cite{ley2}.
    Taking the point of view that the restriction to color
singlet space should cause changes of at most $10-20\%$, but should not
induce radical differences otherwise \cite{bugs}, we attempt to solve the gap 
equations that are derived without this restriction and which have a 
physical constraint imposed upon the quasiparticle energy.   We do this
using the confining potential that was introduced in Section III, following
the analysis of this section rather closely.

We proceed by briefly deriving the finite temperature gap equation.

\subsection{Gap equation}

The Green function approach that was detailed in Section II can be simply
generalized to finite temperatures in order to derive the associated gap
equation.    Here, using the Matsubara or imaginary time propagator,
\begin{equation}
S_\beta(i\omega_n,\vec p) = \frac 1{i\omega_n\gamma_0 - \vec\gamma\cdot
\vec k - \Sigma_\beta(\vec k)},
\label{g1}
\end{equation}
where $\omega_n=(2n+1)\pi/\beta$, $n=0,\pm1,\pm2,\pm3,\dots$, and the 
self-energy $\Sigma_\beta(\vec k)$ is again decomposable into a vector and 
scalar piece, $\Sigma_\beta(\vec k) = \Sigma^S_\beta(\vec k) + \vec
\gamma\cdot
\hat k\Sigma^V_\beta(\vec k)$, one may write
\begin{equation}
\Sigma_\beta(p) = - \frac 1\beta\sum_{a=1}^8\sum_n\int\frac{d^3k}{(2\pi)^3}
V(\vec p - \vec k)\gamma^0 \frac{\lambda^a}2S_\beta(i\omega_n,\vec
k)\gamma_0 \frac{\lambda^a}2,
\label{g3}
\end{equation}
on interpreting the Fock diagram of Fig.2.   Now the steps that were
performed in Section II leading to Eq.(\ref{ep2}) can be similarly executed
here, once the Matsubara sum has been performed.   Eqs.(\ref{ep2}a) and
(\ref{ep2}b) are replaced by their temperature generalizations,
\begin{mathletters}\label{ept}
\begin{eqnarray}
E_\beta(p)\sin(\phi_\beta(p)) &=& \frac23\int\frac{d^3k}{(2\pi)^3}
V(\vec p - \vec k) \sin(\phi_\beta(k)) \tanh(\frac\beta 2 E_\beta(k)) \\
E_\beta(p) \cos(\phi_\beta(p)) - p &=& \frac 23\int\frac{d^3k}{(2\pi)^3}
V(\vec p - \vec k )\hat p \cdot \hat k\cos(\phi_\beta(k))\tanh(\frac\beta 2
E_\beta(k)).
\end{eqnarray}
\end{mathletters}
Note that a central issue in this derivation is that the self-energies
$\Sigma^S_\beta(p)$ and $\Sigma^V_\beta(p)$ making up $\Sigma_\beta(p)$ are now 
necessarily a function of temperature.   This temperature dependence is
now relayed to the reparametrization functions $\phi_\beta(p)$ and
$E_\beta(p)$ that are defined as in Eqs.(\ref{ep}), but which contain the
explicit temperature labels.    It has been noted that the simple procedure
used previously to decouple these equations no longer functions
\cite{ley2,kle}.    The 
finite temperature gap equation, contructed by multiplying Eq.(\ref{ept}a)
by $\cos(\phi_\beta(p))$, Eq.(\ref{ept}b) by $\sin(\phi_\beta(p))$ and
subtracting, is
\begin{eqnarray}
p\sin(\phi_\beta(p)) = \frac 23 \int\frac{d^3k}{(2\pi)^3} V(\vec p - \vec k)
[\sin
(\phi_\beta(k))\cos(\phi_\beta(p)) &-& \hat p \cdot\hat k\cos(\phi_\beta(k))
\sin(\phi_\beta(p))] \nonumber \\
 &\times & \tanh(\frac \beta 2 E_\beta(k)).
\label{g4}
\end{eqnarray}
This gap equation coincides with that of Refs.\cite{kle} and \cite{ley2}.
    It differs
however from that of Ref.\cite{math} in that these authors
regard $E_\beta=E$ as a constant, {\it independent} of the temperature.   In
contrast, in 
Ref.\cite{alk}, the authors use a gap equation similar to Eq.(\ref{g4}), but
with the $\tanh$ factor entirely absent.    These authors claim that the
hyperbolic tangent function and energy dependence is irrelevant, arguing
that $E(k)$ is infinite anyway.    This argument is, however, too naive,
and is not confirmed by our study, in which we first make use of {\it
physical}
arguments to shift the energy along the lines that were discussed in Section
II:
Note that, 
taken at face value, Eq.(\ref{g4}) taken together with Eq.(\ref{ept}a), 
violates the criterion that physical observables, such as $\phi_\beta(k)$
are invariant under a shift in the potential by a constant, as was easily
seen to be the case in Section II at $T=0$.   This issue has been dealt with
in detail in Refs.\cite{kle} and \cite{ley2}.    Both of these sets of
authors show that this conflict can be resolved by supressing color
fluctuations and restricting the grand canonical ensemble trace to cover
color singlet states only.   The color projected gap equation \cite{ley2}
is however extremely difficult to solve, in particular, because of the
temperature dependence of the energy.    Note that the temperature
dependence
of the energy variable is crucial in the BCS theory of superconductivity for 
obtaining the correct temperature behavior of the gap parameter for a
superconductor \cite{fetter}.    It is thus the
purpose of this study to investigate the role of the temperature dependence
of the energy function and its consequences.     

Guided by the situation at $T=0$, one may make a simple physical {\it
ansatz},
 that of replacing
$E_\beta(p)$ in the $\tanh$ function of Eq.(\ref{g4}) by
\begin{equation}
\bar E_\beta(p) = E_\beta(p) - E_\beta(0) + m^*_\beta.
\label{g5}
\end{equation}
This  is motivated, as in Section II following Ref.\cite{adler}, by
comparing the Green function in this case with the low momentum version of
the free temperature Green function \cite{kle}.    The real time forms are
\begin{eqnarray}
S(\vec p&,&\omega)= [
\frac{R_{+,\beta}(p)}{\omega - E_\beta(p) + i\eta} + 
\frac{R_{-,\beta}(p)}{\omega + E_\beta(p) - i\eta}] \nonumber \\
& + & 2\pi i f[\beta E_\beta(p)][\gamma^0\omega - \vec \gamma\cdot \hat p
E_\beta(p)\cos (\phi_\beta(p)) + E_\beta(p)\sin(\phi_\beta(p))]\delta(
\omega^2 - E^2_\beta(p)), \nonumber \\
\label{g6}
\end{eqnarray}
for the full Green function, with 
\begin{equation}
R_{\pm,\beta}(p) = \frac 12[\gamma^0\mp\vec \gamma\cdot \hat p
\cos(\phi_\beta(p)) \pm\sin(\phi_\beta(p))],
\label{g7}
\end{equation}
where
$f(x) = 1/(\exp(\beta x) + 1)$ and $\eta\rightarrow 0^+$, while the free
Green function has the form
\begin{eqnarray}
S_{free} &=& \frac{R^0_{+,\beta}(p)}{\omega-m^*_\beta + i\eta} + 
\frac{R^0_{-,\beta}(p)}{\omega + m^*_\beta - i\eta} \nonumber \\
&+& 2\pi i f(\beta m^*_\beta)[\gamma_0\omega-\vec \gamma\cdot \vec p +
m^*_\beta] 
\delta(\omega^2 - m_\beta^2) 
\label{g8}
\end{eqnarray}
with $R^0_{\pm,\beta}(p) = \frac12(\gamma^0\mp\vec \gamma\cdot\vec p/m^*_\beta
\pm1).$

Thus one recovers the identifying criterion
\begin{equation}
\lim_{p\rightarrow 0}\frac {\cos(\phi_\beta(p))} p = \frac 1{m^*_\beta}
\label{g9}
\end{equation}
that generalizes Eq.(\ref{mass}) to finite temperatures, as well as the
essential condition
\begin{equation}
E_\beta(p)\rightarrow \bar E_\beta(p)\rightarrow m^*_\beta, \quad p\rightarrow 0
\label{g10}
\end{equation}
that has prompted the replacement Eq.(\ref{g5}).   Given Eq.(\ref{g5}),
one recovers the expected physical quasiparticle properties.    

The associated finite temperature condensate density is determined to be
\begin{equation}
\langle\bar\psi\psi\rangle_\beta= -\frac 3{\pi^2}\int_0^\infty dk
k^2\sin(\phi_\beta(k))\tanh(\frac\beta2\bar E_\beta(k)),
\label{g11}
\end{equation}
which will be evaluated in the following section for the potential for  quadratic
confinement.

\subsection{Quadratic Confinement}

The purpose of this section is to examine the temperature dependence of
the gap function and the self-energies for the quadratically 
 confining potential of
Eq.(\ref{quad}) and its Fourier transform, Eq.(\ref{pot}).
In this section, the strategy is to implement Eq.(\ref{g5})  in an
approximate fashion that highlights the temperature dependence of the
quasiparticle energy.   We proceed once again by bringing
the gap equation into differential form.    One finds that
Eq.(\ref{difge}) has the generalized form
\begin{eqnarray}
ap^3\sin(\phi_\beta(p))\coth(\frac{\beta}2\bar E_\beta(p)) &=& p^2
\phi'{}'{}_\beta(p) + 2p\phi_\beta'(p) + \sin(2\phi_\beta(p)) \nonumber \\
&+& 2p^2\frac{\beta\bar E_\beta{}'(p)}{\sinh(\beta\bar E_\beta(p))}\phi_
\beta{}'(p),
\label{horror}
\end{eqnarray}
while the energy equation, Eq.(\ref{en}) is generalized to read
\begin{eqnarray}
\bar E_\beta(p) &=& -\frac 1 a(\phi_\beta{}'(p)^2 + \frac
2{p^2}\cos^2(\phi_\beta(p)))\tanh(\frac{\beta}2\bar E_\beta(p)) + p\cos(
\phi_\beta(p)) \nonumber \\
 &+& \frac 1{ap}[p\tanh(\frac{\beta}2\bar E_\beta(p))]'{}' + 
\frac23U\tanh(\frac\beta 2\bar E_\beta(p)).
\label{horrore}
\end{eqnarray}
Unlike the system of equations (\ref{difge}) and (\ref{en}) at zero
temperature, one sees that Eqs.(\ref{horror}) and (\ref{horrore}) are coupled non-linear
differential equations in the two variables $\phi_\beta$ and $\bar E_\beta$.
While it is possible to make a simple analysis for high temperatures and
low/high momenta, it is a difficult problem to solve these two equations
exactly.    We thus do not attempt to do so here.  After discussing the limits
just mentioned, we shall investigate the role of an {\it assumed} temperature
dependent dispersion relation for the energy, that allows us to decouple
these equations and solve them.

\paragraph{High temperature limit.}   If, as $T\rightarrow\infty$, $\bar
E_\beta(p)$ remains finite, then from Eq.(\ref{horror}),
\begin{equation}
p\sin(\phi_{\beta\rightarrow 0}(p)) = 0  \quad\mbox{for all} \ p
\label{t1}
\end{equation}
so that
\begin{equation}
\phi_{\beta\rightarrow0} (p) = 0.
\label{t2}
\end{equation}
Consequently, from Eq.(\ref{horrore}), it follows that
$\bar E_{\beta\rightarrow0}=p$ for all $p$, and one recovers the chiral limit.

\paragraph{Small momenta.}  Using the boundary conditions, $\bar E_\beta(0)
=m_\beta$ and $\phi_\beta(0) = \pi/2$, one can expand $\phi_\beta(p)$ and
$\bar E_\beta(p)$ in the small momentum region,
\begin{mathletters}\label{expand}
\begin{eqnarray}
\phi_\beta(p) &=& \frac \pi 2 + c_1^\beta p + c_2^\beta p^2 + \dots  \\
\bar E_\beta(p) &=& m_\beta + a_1^\beta p + a_2^\beta p^2 + \dots 
\end{eqnarray}
\end{mathletters}
where, from Eq.(\ref{g9}),
\begin{equation}
c_1^\beta = -\frac 1{m^*_\beta} = c_\beta.
\label{t3}
\end{equation}
One may insert these expansions into Eq.(\ref{horrore}).   A necessary
condition that $\bar E_\beta(p)$ does not diverge as $p\rightarrow 0 $ is
that $a_1^\beta = 0$.   Thus one regains the observed $T
\rightarrow 0$ dependence of Section III that
\begin{equation}
\bar E_\beta(p) = m^*_\beta + a_2^\beta p^2, \quad \mbox{for } \ p\rightarrow
0,
\label{t4}
\end{equation}
this equation being the analog of Eq.(\ref{dis}).   As in that case, one may
continue the expansion for $\phi_\beta(p)$, writing
\begin{equation}
\phi_\beta(p) = \frac \pi 2 + c^\beta p + \eta_\beta(p), \quad
\mbox{with}\  \eta_\beta(0)=0.
\label{t5}
\end{equation}
Inserting this and Eq.(\ref{expand}b) into the gap equation enables one to 
find a differential equation for $\eta_\beta(p)$ which has the solution
$\eta_\beta(p) = b^\beta p^3$, with
\begin{equation}
b^\beta = \frac 1{10}[a\coth(\beta/2|c^\beta|) +\frac{4\beta a_2^\beta
|c^\beta|}{\sinh(\beta/|c^\beta|)} + \frac 4 3(|c^\beta|)^3].
\label{t6}
\end{equation}
The coefficient $a_2^\beta$ of the energy expansion may also be determined
by inserting the expansions for $\bar E$ and $ \phi$ into
Eq.(\ref{horrore}).      One finds
\begin{equation}
a_2^\beta = \frac 13[3|c^\beta|^2\tanh(\beta/2|c^\beta|) + a(\frac
1{|c^\beta|} - \frac 23 U\tanh(\beta/2|c^\beta|))]\frac 1\beta\cosh^2(\beta
/2|c^\beta|).
\label{t7}
\end{equation}
Thus the behavior of $\phi_\beta$ and $\bar E_\beta$ for small momenta is
similar to that obtained at $T=0$, see Section III.

\paragraph{Large momenta.}   From Eq.(\ref{horrore}), it follows that, at
large momenta,
 $\bar E_\beta(p) \sim
p$, $p\rightarrow \infty$, and one no longer has a temperature dependence
in this quantity.  Thus  the gap equation is precisely that of the zero
temperature case.

As has been seen at zero temperature, and now also at finite values of the
temperature, the dispersion relation for the energy is non-relativistic in
the low momentum regime, but relativistic for high momenta.   This is
commensurate with a general relativistic {\it ansatz} for the quasiparticle
energy.   In the following, we will discuss the solution to the gap equation
for two possible {\it ans\"atze}:  (i)  momentum independent, but temperature 
dependent,
\begin{equation}
\bar E_\beta(p) = m^*_\beta
\label{a1}
\end{equation}
and (ii) the relativistic dispersion relation,
\begin{equation}
\bar E_\beta(p) = \sqrt{m^*_\beta{}^2 + p^2}.
\label{a2}
\end{equation}
Note that {\it ansatz} (i), being totally devoid of momentum dependence, 
serves to highlight the influence of the {\it temperature} dependence of
the energy variable on determining the features of the solution to the 
gap equation.   In addition, the simplicity of this {\it ansatz} allows us
to solve the resulting equations easily.     The second
{\it ansatz}, while meeting physical expectations more readily, leads 
to equations that are somewhat more difficult to solve.   As shall become
evident, the results obtained from the second {\it ansatz} do not differ
qualitatively from that of {\it ansatz} (i).     This appears to indicate
that the momentum dependence is secondary in comparison to a correct
treatment of the implicit temperature behavior of the quasiparticle energy.
Our  two choices are detailed in the following two subsections.

\subsection{Momentum independent ansatz}

In this subsection, we examine the solution of the temperature dependent
gap equation under the {\it ansatz} Eq.(\ref{a1}).    This is interesting to 
study since it contains the essential ingredient, i.e. temperature
dependence, that has previously been overlooked.   Furthermore, the
resulting transcendental equation is simple to solve.   Let us be more
precise.    Inserting  Eq.(\ref{a1}) into Eq.(\ref{horror}) leads to the
result
\begin{equation}
a^\beta p^3\sin(\phi_\beta(p)) = p^2 \phi_\beta'{}'(p) + 2p\phi_\beta(p) +
\sin(2\phi_\beta(p)),
\label{gea1}
\end{equation}
with
\begin{equation}
a^\beta = a\coth(\frac \beta 2 m_\beta^*).
\label{gea12}
\end{equation}
Except for the cotangent factor, this equation is identical with the
$T=0$ equation,   Eq.(\ref{difge}).    It may thus be treated
semi-analytically in the same fashion.   In particular, the relation
Eq.(\ref{tt3})  is valid, with $c$ replaced by $c^\beta$ and $a$
replaced by $a^\beta$, i.e.
\begin{equation}
|c^\beta| = \frac{(a^\beta)^{1/3}}3\exp(\frac2{\sqrt 7}(n\pi - \rho-\delta)).
\label{a3}
\end{equation}
Now, inserting $a^\beta$ from Eq.(\ref{gea12}) into this equation leads to
the
simple result
\begin{equation}
|c^\beta| = (\coth(\frac\beta 2 m_\beta^*))^{1/3}|c|,
\label{a4}
\end{equation}
which, together with identification $|c^\beta| = 1/m_\beta^*$ and
$|c|=1/m^*$, leads to the transcendental equation
\begin{equation}
m_\beta^* = m^*(\tanh(\frac\beta 2 m_\beta^*))^{1/3}
\label{a5}
\end{equation}
that determines $m_\beta^*$.   In addition to the trivial solution, $
m_\beta^*=0$, this equation has a non-trivial solution that can be 
determined numerically.   Following this, $\phi_\beta(p)$ is determined, and
the condensate $\langle\bar\psi\psi\rangle$, from Eq.(\ref{g11}), can
be shown to have the form
\begin{equation}
\langle\bar\psi\psi\rangle_\beta = \langle\bar\psi\psi\rangle_{T=0}
(\tanh(\frac\beta 2m_\beta^*))^2
\label{a6}
\end{equation}
in relation to its zero temperature value.

In Fig.4, we show the values of $\phi_\beta(p)$ as a function of $p$ for 
different values of the temperature, scaled with respect to the dynamically
generated constituent quark mass.   One notes from this figure that the 
slope increases with increasing temperature, indicating the fall off of the
constituent quark mass.   The temperature dependence of the condensate 
is shown in Fig. 5 as a function of the dimensionless variable $T/m^*$.
One sees that it falls off strongly to zero as $T\rightarrow0$:
at $T/m^*=2.0$, the value of the condensate is less than 5\% of its value
at the origin.    However, the transition of the condensate is smooth and
there is no evident critical temperature at which a phase transition that 
restores chiral symmetry occurs.    This behavior is also evident in that of
the quasiparticle mass, which is shown in Fig.5 as a function of $T/m^*$.
This function decreases rapidly over the first unit of $T/m^*$, but
thereafter drifts gradually to zero.

\subsection{Relativistic dispersion relation}

In this subsection, we investigate the use of a relativistic energy-momentum
{\it ansatz}, viz. that of Eq.(\ref{a2}).   For small values of $p$, this has
the expansion
\begin{equation}
\bar E_\beta(p) \stackrel{p\rightarrow 0}{=} m_\beta^* +
\frac{p^2}{2m_\beta^*} + O(p^4),
\label{ex}
\end{equation}
enabling one to identify the expansion coefficient of the energy as
\begin{equation}
a_2^\beta = 
\frac 1{2m_\beta^*} = \frac 12 |c^\beta|,
\label{r2}
\end{equation}
and that of the gap parameter, from Eq.(\ref{t6}).   One then has
\begin{equation}
b^\beta = \frac 1{10} [a\coth(\frac \beta{2|c^\beta|}
) + \frac {2\beta
|c^\beta|^2}{\sinh(\beta/|c|^\beta)} + \frac 4 3|c^\beta|^3].
\end{equation}
The estimate for the gap parameter, $\phi_\beta(p)$, for $p$ small, thus
enables one to perform a numerical integration starting at a value of $p$
close to the origin.   Fig. 6 shows the solution to the gap equation as a
function of momentum at different values of the temperature.    This bears
close resemblance to the solution obtained in the previous case.   The 
condensate, calculated numerically from Eq.(\ref{g11}), is shown in Fig.~7.
As in the previous case, it goes to zero, but somewhat faster than in the 
previous case.   Once again, no transition to a chirally symmetric phase is
observed at a finite value of the temperature, although the fall off, for
low temperatures, is steep.

\subsection{Comparative analysis}

It is useful to examine the results for the previous two calculations on one
graph.   In Fig.~8, we show the condensate as a function of the temperature, 
for two arbitrary choices of the mass at zero temperature, either
$m^*=200$MeV, or $m^*=300$MeV.  It is interesting to note that both 
assumptions for the dispersion relation lead to the same qualitative
picture.   Both curves fall rapidly initially.  Instead of a phase
transition
occurring however, the curves flatten off and fall more gently to zero.
One may speculate that a better form or exact solution of the coupled
equations may in fact yield a chiral phase transition at a finite value
of the temperature.   In any event, it is heuristically evident that the
temperature fluctuations alone could drive a phase transition.   In view
of the difficulties presented here, we must conclude that it is necessary to study
temperature behavior of even simpler confining models.

The temperature behavior of the masses themselves is shown in Fig.~9.
These are also decreasing with temperature, as occurs, for example, in the
Nambu--Jona-Lasinio model, but no direct linear proportionality with the 
condensate is in evidence.     The fact, however, that the quasiparticle 
masses are a function of temperature is an indication that this feature,
which gives rise to dropping or rising composite particle masses, will
continue to be an existing feature of a model that incorporates confinement
and chiral symmetry and which one can hopefully solve exactly.   It is to
be believed that the solutions obtained here, which in both cases indicate
a drop in the condensate energy, but no phase transition at finite
temperature, are a consequence of the approximations that have been made.

\section{SUMMARY AND CONCLUSIONS}

In this paper, we have analysed the Coulomb gauge model with respect to the 
inclusion of a divergent potential.   In this model, confinement is
incorporated into the scalar sector of the vector four-potential, although
there is no compelling reason why this should be done.   The
consequences for the quark propagator are discussed and an analysis of the
divergences in the infrared sector as well as the ultraviolet is presented.
The arguments of Adler and Davis \cite{adler}
 that physical quantities should be 
infrared finite, coinciding with the condition that a shift in the potential
by a constant should not affect physical quantities for color singlet states
is used as the determining criterion for physical observables.    An
interpretation of the quasiparticle mass is made, since the quasiparticle 
energy is necessarily divergent.

The finite temperature generalization of this model yields a gap equation
which differs from that of several authors.   The identification of the
quasiparticle mass in the fashion that was done at $T=0$ leads to a natural
physical prescription that suggests the introduction of a non-divergent
temperature dependent quasiparticle energy $\bar E_\beta(p) = E(p) - E(0) + 
m_\beta^*$, so that $\bar E_\beta(p)\rightarrow m_\beta^*$ as $p\rightarrow
0$.    Even so, the complex set of coupled equations that one obtains for
the
gap parameter and the energy cannot be solved exactly.    Since the issue of 
other authors has been to neglect the temperature dependence of the energy 
function and obtain alternatively that a phase transition can \cite{alk} or cannot
\cite{ley2} occur, we propose to investigate two simple dispersion forms for the
quasiparticle energy that have an inherent temperature dependence.    We
find
that, while there is no transition to a chirally symmetric phase at a finite
value of the temperature, there is nonetheless a sharp drop in the
 condensate density or order parameter at some point.   Whether this is 
endemic to a confining potential, or whether it is simply a consequence of
the shortcomings of the approximations made, is as yet unknown and could
be studied using other confining potentials, and simplifying models.   
It is perhaps therefore of interest to develop models that are closer in
structure to that of Nambu and Jona-Lasinio, in order to understand these
properties.

\section*{ACKNOWLEDGMENTS}

We wish to thank P. Rehberg and J. H\"ufner   for their criticisms and careful reading of the 
manuscript.
This work has been supported in part by the Deutsche Forschungsgemeinschaft
DFG under the contract number Hu 233/4-4, and by the German Ministry for
Education and Research (BMBF) under contract number 06 HD 742.

\eject
TABLE I

\vskip 0.4in

\begin{tabular}{|| c | c | c ||} \hline
n  & $|c_{asmpt}|$ & $|c_{num}|$  \\ \hline
0  & 2.09169       & 2.03750      \\
1  & 22.4837       & 22.4246      \\
2  & 241.680       & 241.634      \\
3  & 2597.84       & 2597.83      \\
4  & 27924.4       & 27925.0      \\
5  & 300161        & 300170       \\
\hline
\end{tabular}

\vskip 0.5in
Table 1: Solutions for the slope $c$ obtained via asymptotic matching of 
the approximate solutions, and via a numerical procedure.   The symbol
$n$ denotes the number of nodes that the solution has.

\eject

FIGURE CAPTIONS

\vskip 0.2in

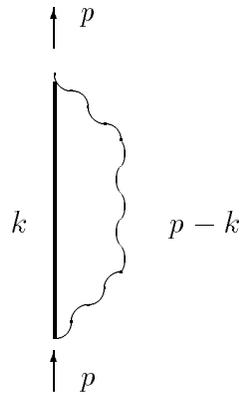
Fig.~1.   Fock Feynman diagram for the self-energy.

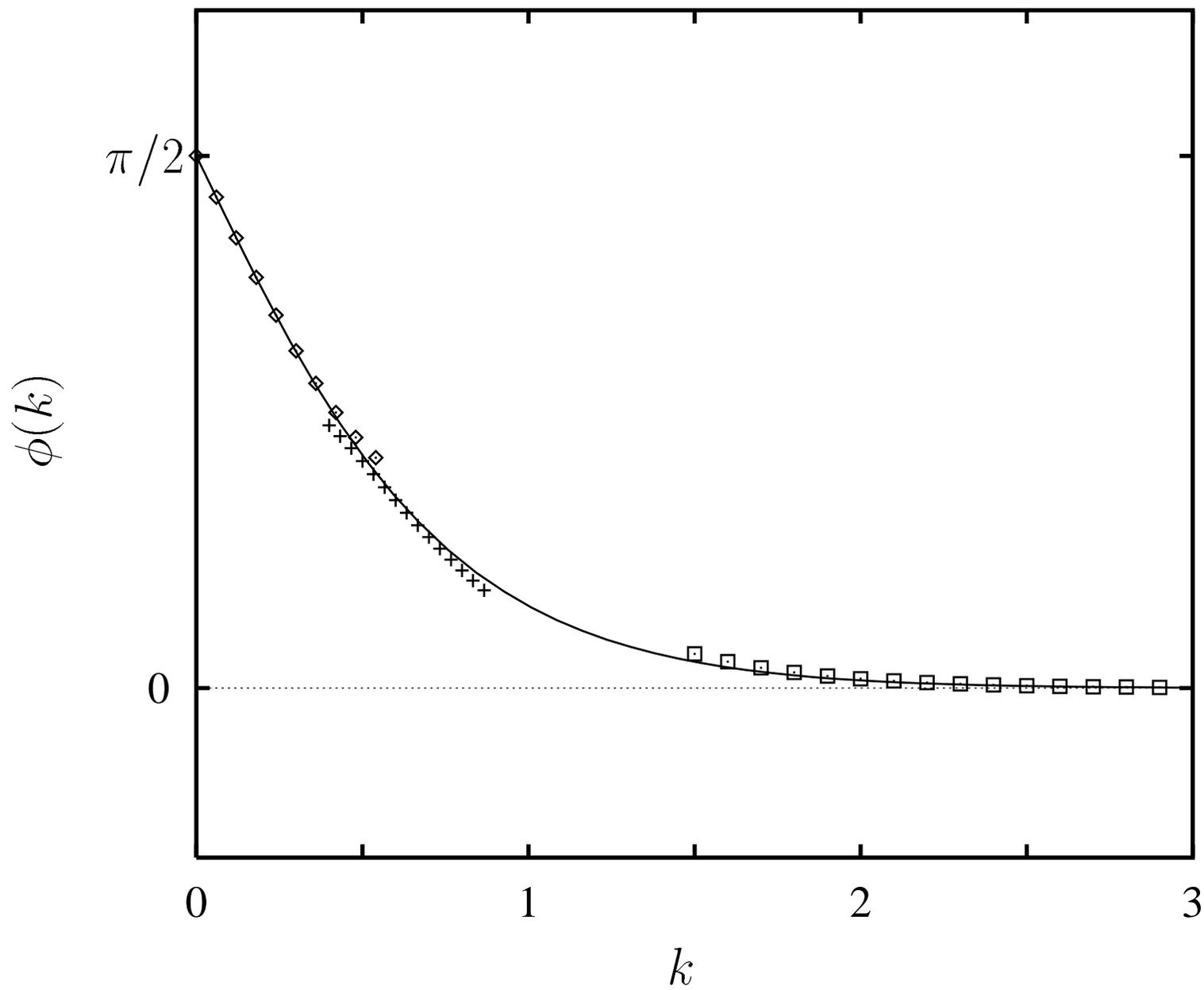
Fig.~2.   The nodeless solution for the gap angular variable is shown
as a function of the scaled momentum $k=a^{1/3}p$.    The semi-analytic solutions in regions
(1), (2) and (3) are indicated by crosses, stars and circles respectively.

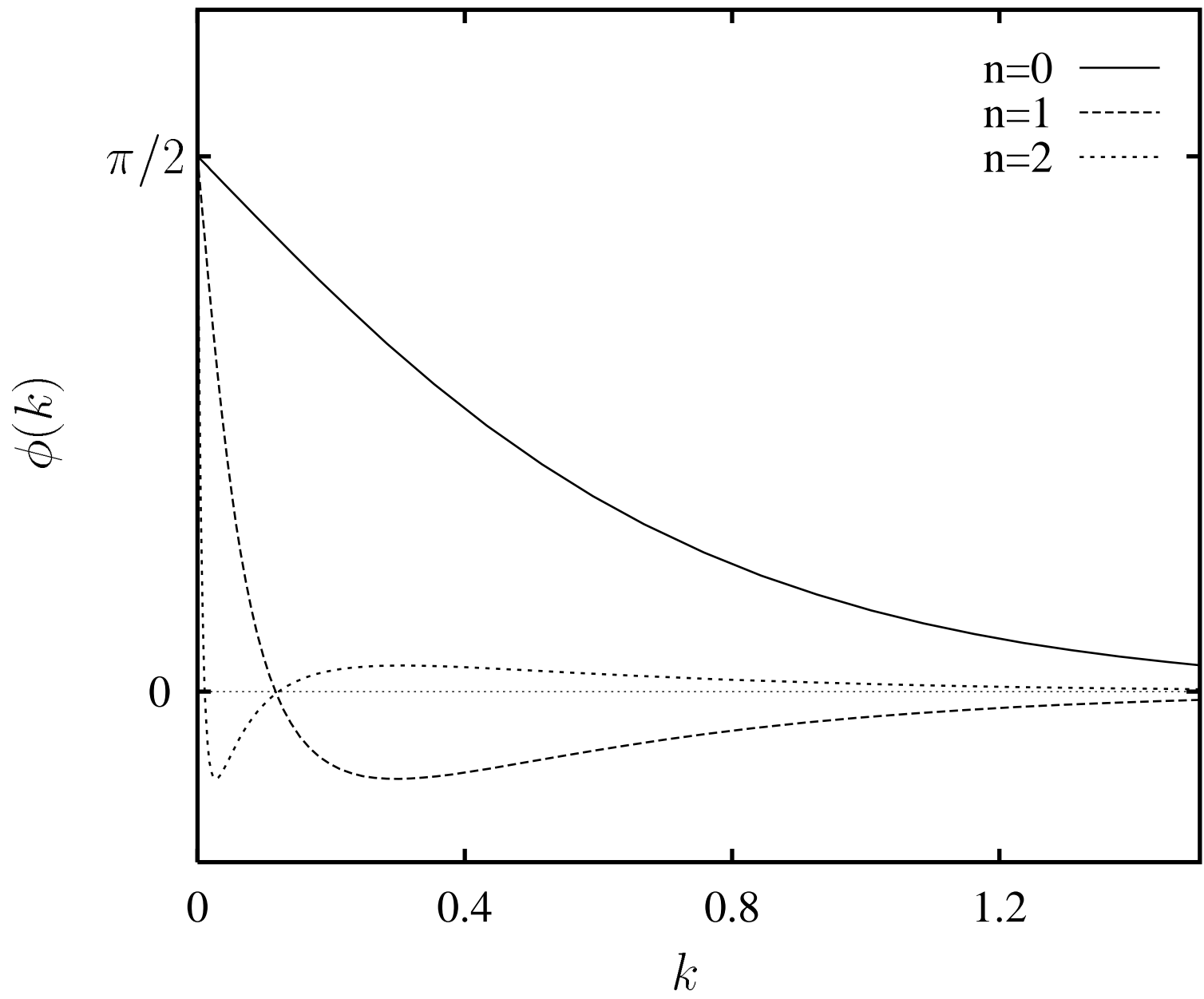
Fig.~3.   The first three solutions for $\phi(k)$ with zero, one and two
nodes, are shown as a function of
the scaled momentum $k=a^{1/3}p$.

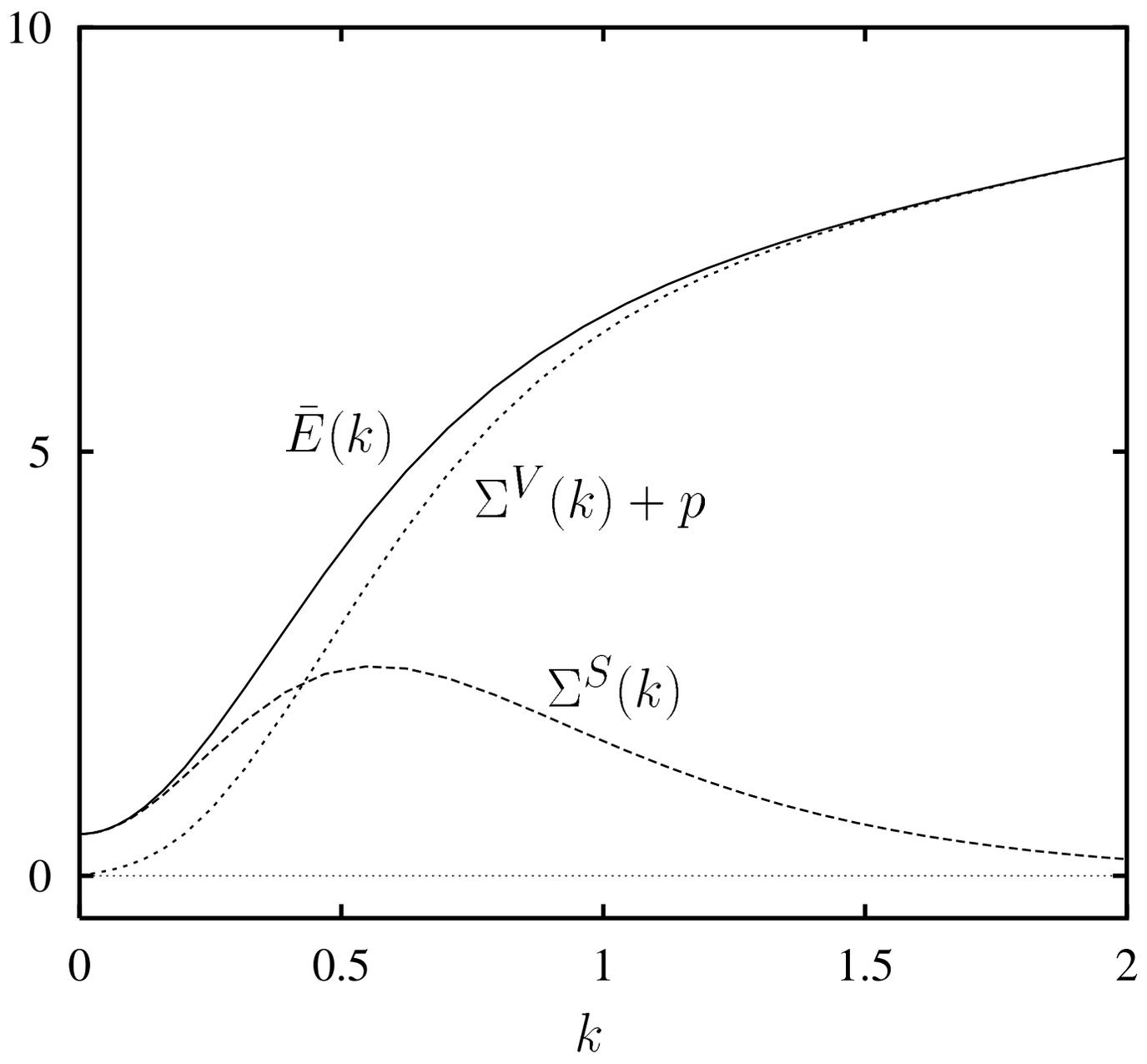
Fig.~4.  The quasiparticle spectrum $\bar E_\beta(k)$ and the scalar and 
vector self-energies $\Sigma^S(k)$ and $\Sigma^V(k) + k$ are shown as a
function of the scaled momentum $k=a^{1/3}p$.

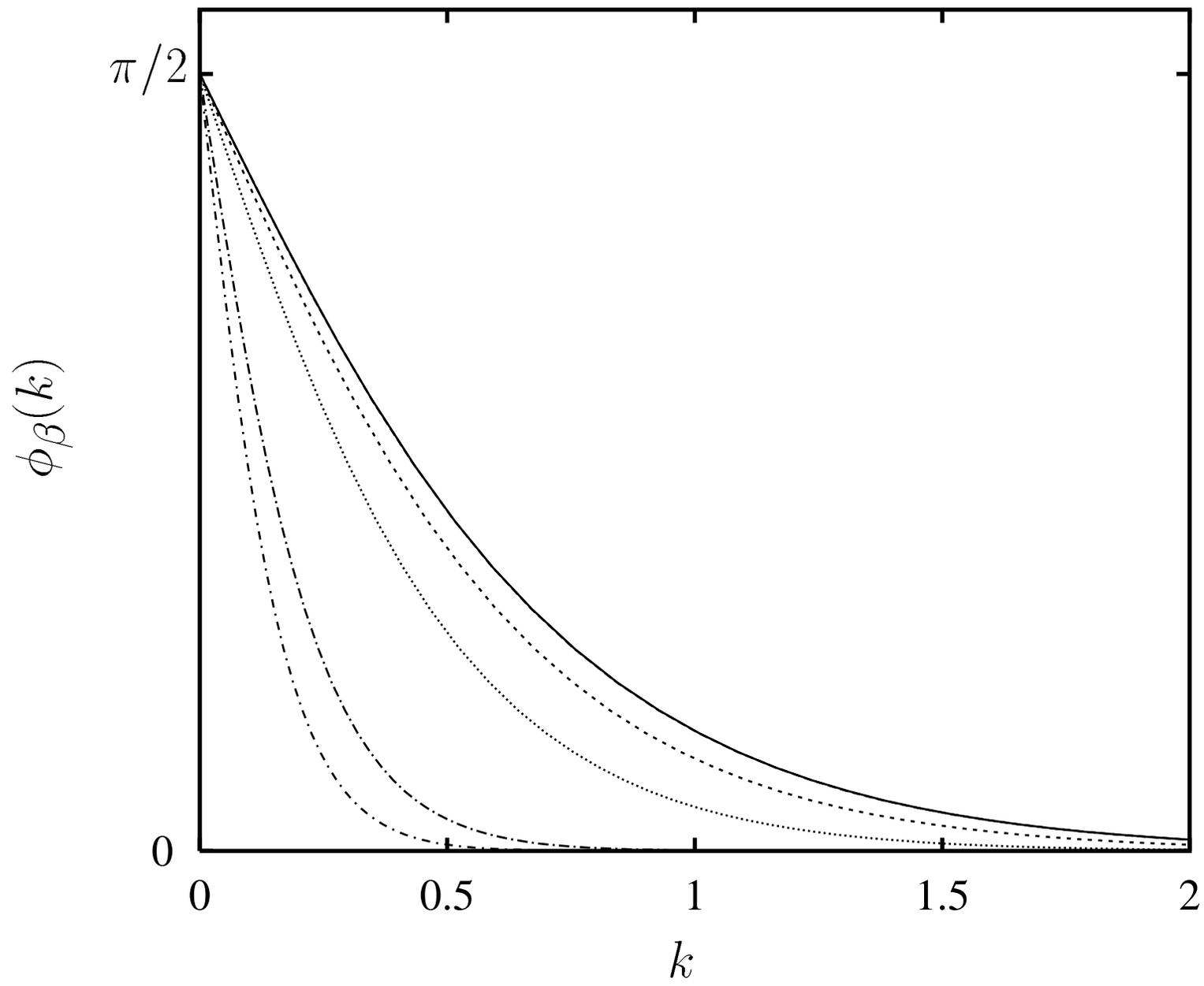
Fig.~5. Solutions $\phi_\beta(p)$ as a function of $p$, in the approximation
$\bar E_\beta=m_\beta^*$ for various values of the scaled temperature
$T/m^*$.   The upper curve corresponds to $T/m^*=0$.   Successive curves
have the values $T/m^* = 0.2, 0.5,1.0,5.0$ and $10.0$.

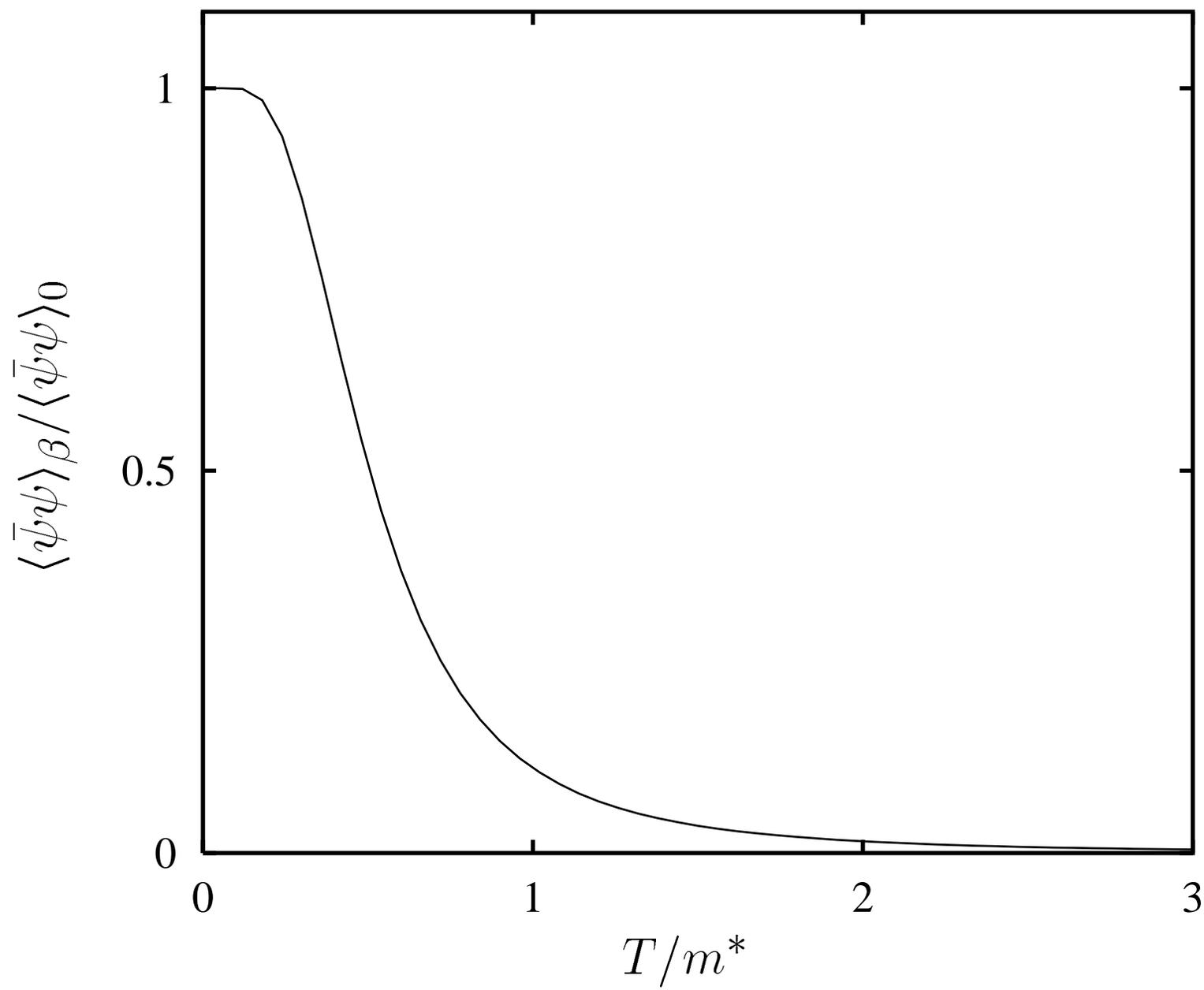
Fig.~6.  The temperature dependence of the condensate, normalized to its
zero temperature value.  The approximation $\bar E_\beta(p) = m_\beta^*$
has been used.

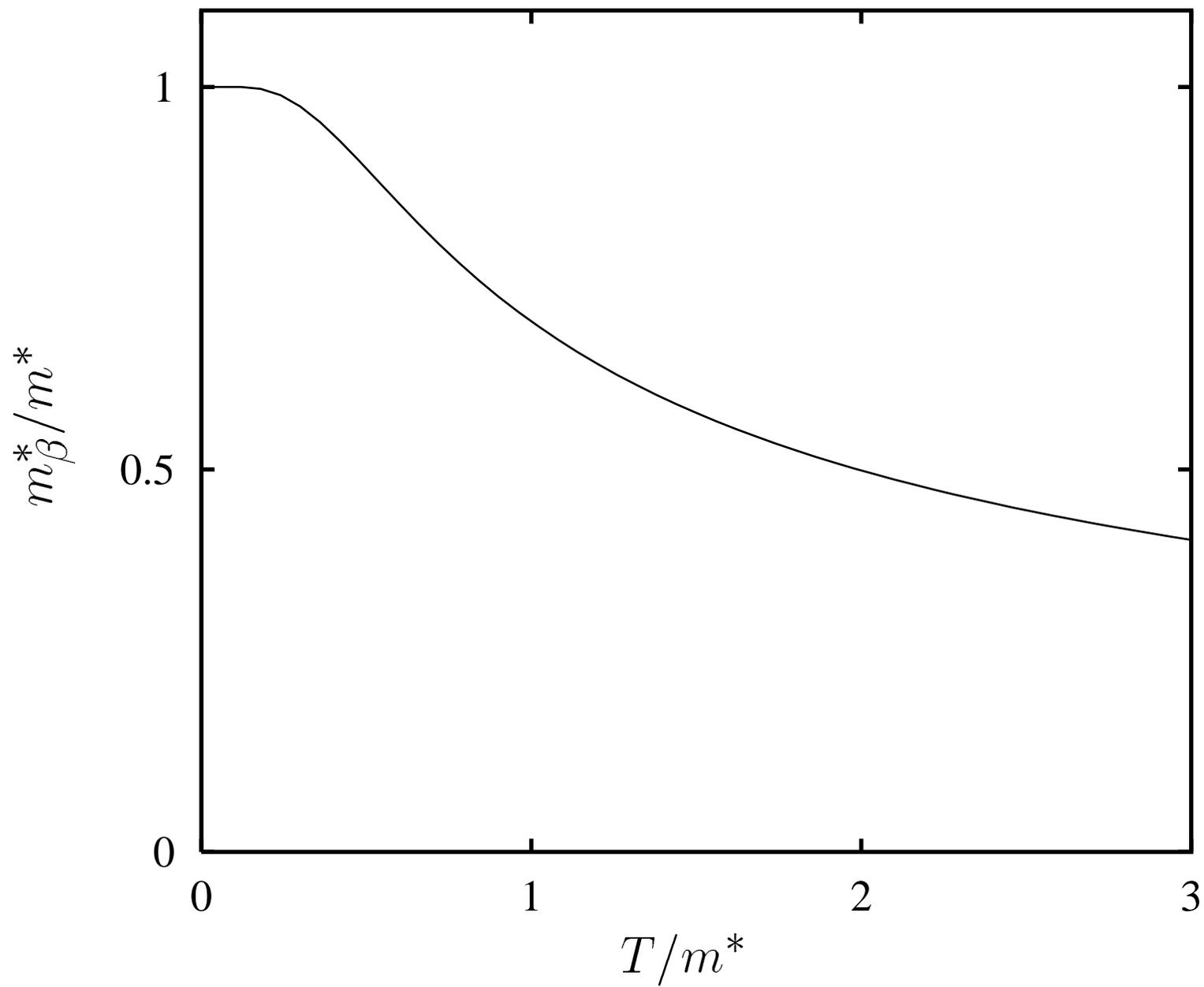
Fig.~7.  The temperature dependence of the dynamically generated quark mass
is given as a function of the scaled temperature $T/m^*$.

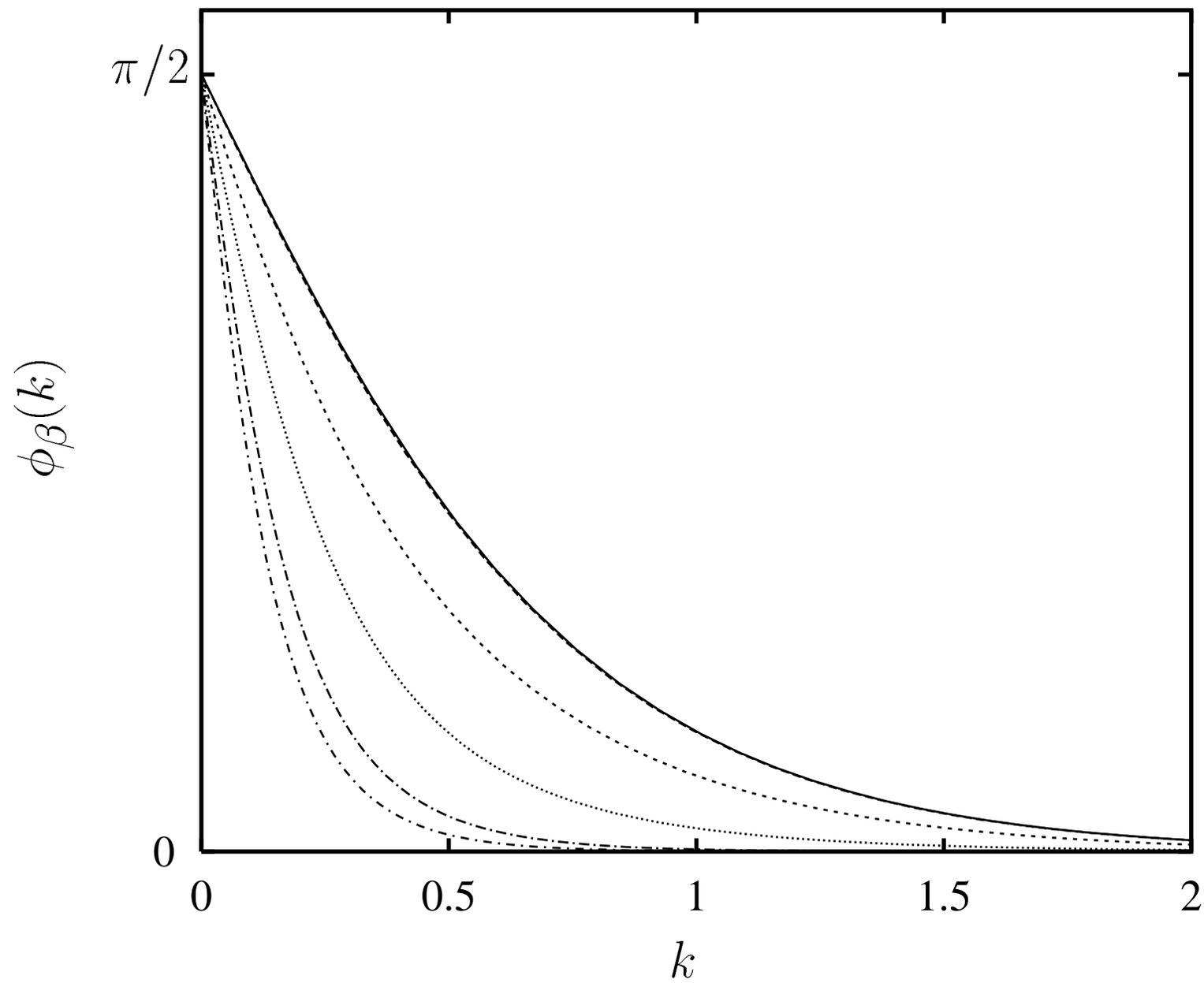
Fig.~8.  As in Fig.~5, with the momentum dependent ansatz $\bar
E_\beta^2(p) = m^*_\beta{}^2 + p^2$.

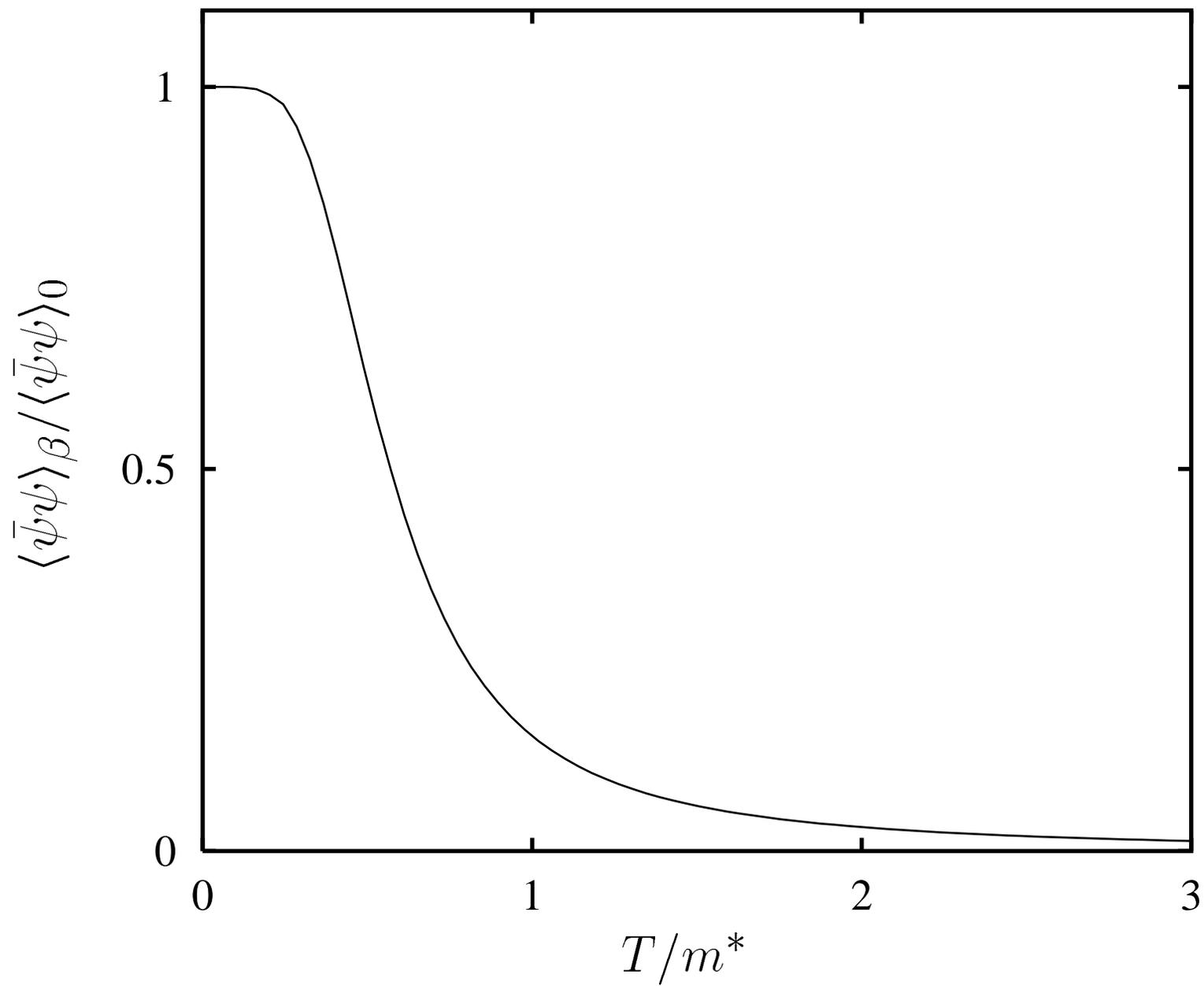
Fig.~9.  As in Fig.~6, with the momentum dependent ansatz $\bar E_\beta^2
(p) = m^*_\beta{}^2 + p^2$.

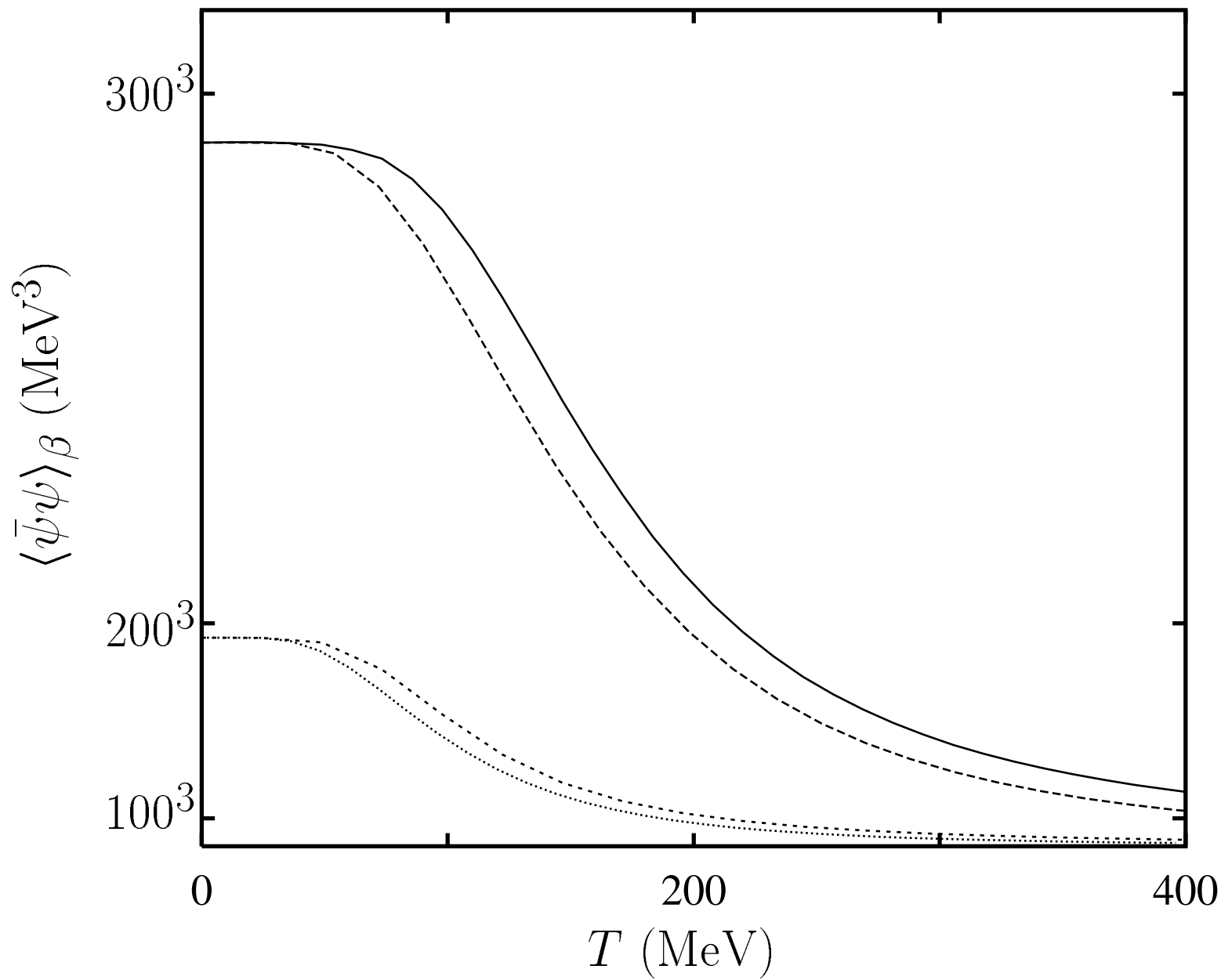
Fig.~10.  The condensate, shown for two arbitrary choices of the dynamically generated
masses $m^*= 200$MeV and $m^*=300$MeV.  The upper curves in each case 
correspond to the momentum independent ansatz $\bar E_\beta(p) =m_\beta^*$,
while the lower correspond in each case to the dispersion ansatz
$\bar E_\beta^2(p) = p^2 + m_\beta^*{}^2$

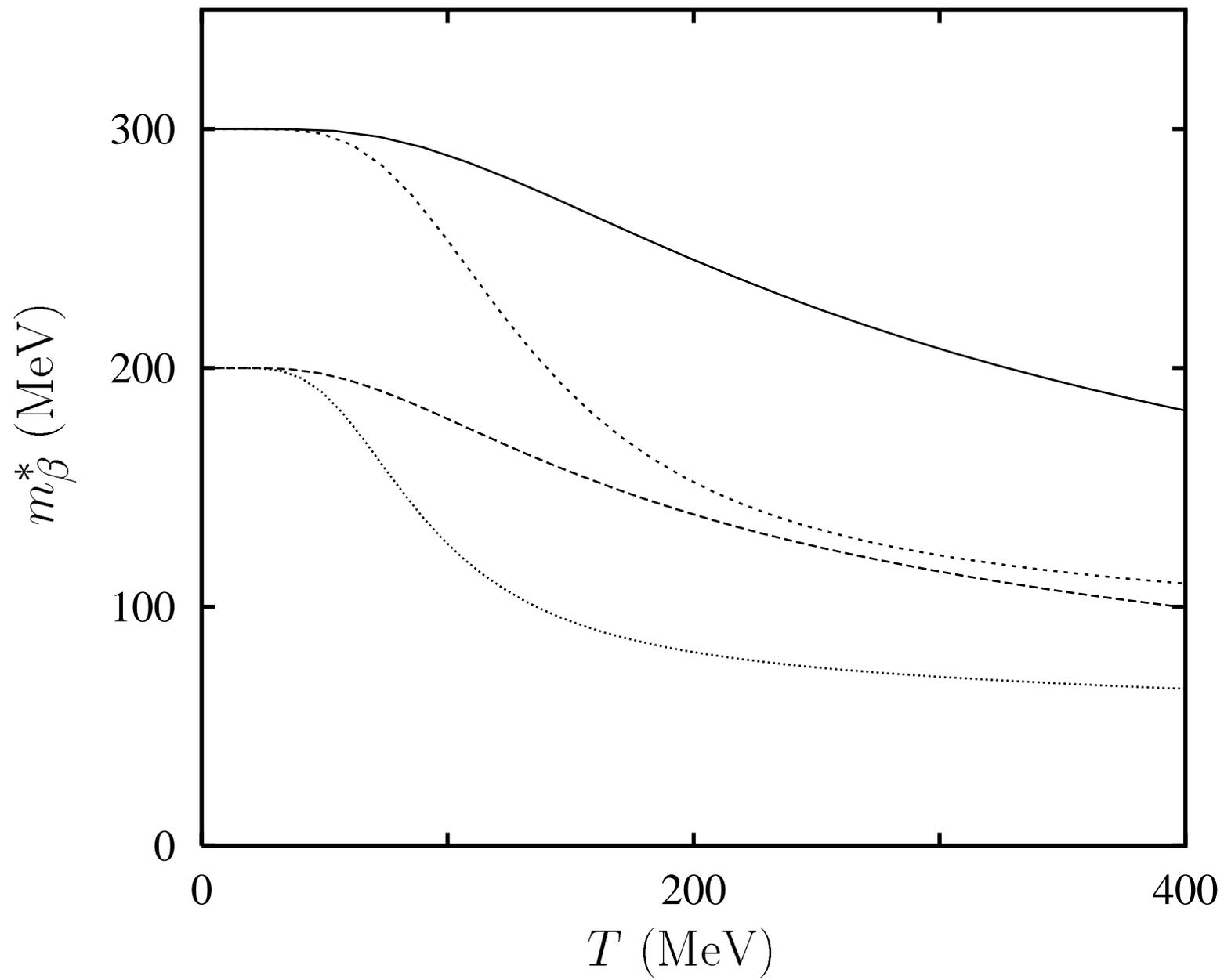
Fig.~11.   The current quark masses, shown for two arbitrary choices of the 
dynamically generated masses $m^*=200$MeV and $m^*=300$MeV.    The upper 
curves in each case correspond to the momentum independent ansatz
$\bar E_\beta(p) = m_\beta^*$, while the lower correspond to the dispersion
ansatz $\bar E_\beta^*(p) = p^2 + m_\beta^*{}^2$. 

\eject


\end{document}